\documentclass[onecolumn,10pt]{revtex4}
\usepackage{color}
\usepackage{bm}
\usepackage{graphics,epsfig}
\usepackage{amsmath}
\usepackage{amssymb,amsfonts}
\usepackage{setspace}
\usepackage{hyperref}
\usepackage[a4paper, top=1.4in, bottom=1.4in, left=1.5in, right=1.5in]{geometry}

\begin{document}

\newtheorem{theorem}{Theorem}
\newtheorem{proposition}[theorem]{Proposition}
\newtheorem{corollary}[theorem]{Corollary}
\newtheorem{example}{Example}

\title{\small  QUANTUM PHASE SPACE, QUANTIZATION HIERARCHY, AND ECLECTIC QUANTUM MANY-BODY SYSTEM  }
\author{Dong-Sheng Wang}
\thanks{\texttt{wdscultan@gmail.com} \;Institute for Quantum Science and Technology, Department of Physics and Astronomy,
University of Calgary, Alberta T2N 1N4, Canada}

\date{\today}

\begin{abstract}
An operator-valued quantum phase space formula is constructed.
The phase space formula of Quantum Mechanics
provides a natural link between first and second quantization,
thus contributing to the understanding of quantization problem.
By the combination of {\em quantization} and {\em hamiltonization} of dynamics,
a {\em quantization hierarchy} is introduced,
beyond the framework of first and second quantization and
generalizing the standard quantum theory.
We apply our quantization method to quantum many-body system
and propose an {\em eclectic model}, in which
the dimension of Hilbert space does not scale exponentially with
the number of particles due to the locality of interaction,
and the evolution is a constrained Hamiltonian dynamics.
\end{abstract}

\maketitle
\tableofcontents
\begin{spacing}{1.2}

\section{Background}

Quantization problem is at the heart of
the mathematical foundation of quantum theory~\cite{TE05}
as well as a major obstacle for a satisfactory interpretation of
quantum objects such as the wavefunction~\cite{Weinb}.
Although different quantization methodologies have been developed~\cite{TE05},
such as geometric quantization, deformation quantization etc,
there is still no unified {\em the} quantization method.
To quantize a classical system, for instance,
it essentially requires mapping functions to operators,
for which there is no unique and consistent way.

Along with the quantization problem,
there exists the division of first quantization and second quantization.
While the first quantization is well understood as
the quantization of classical systems,
there are some ambiguities of the second quantization,
which may refer to the quantization of fields
(e.g., classical scale field, Dirac field, etc),
or the quantization of some first-quantized systems.
The problem of second quantization
is also closely related to the problem of the origin of
the Boson/Fermion statistics,
for which there are controversies about
whether it is a result of second quantization or Special Relativity~\cite{MR03}.

To study the quantization problem,
it is crucial to realize that
classical system is often described in phase space,
while quantum system is often described in Hilbert space~\cite{Dirac,Neuman}.
For many-body system,
the dimension of phase space grows linearly with particle number,
while the dimension of Hilbert space grows exponentially,
which in turn forms the basic motivation of quantum computation~\cite{NC00,Fey82}.
However, there exists a peculiar connection between Hilbert space and phase space formula,
which is revealed by the Geometric Quantum Mechanics approach~\cite{Kibb79,CMP90,CCM07}.
In this approach, the Schr\"odinger equation can be written as Hamilton's equations
associated with a phase space,
and the Hamiltonian function is the expectation value of energy
$H=\langle \psi | \hat{H}|\psi \rangle$.

In this work, we follow the methodology of Geometric Quantum Mechanics
and develop the framework of {\em operator-valued quantum phase space}.
There are some reasons for such a name:
first, in order to distinguish it from the `quantum phase space'
in the geometric approach~\cite{Kibb79,CMP90,CCM07}
which is not operator-valued;
and second, to distinguish it from the `quantum phase space'
based on Wigner function and classical variables (`position' and `momentum'),
which is often employed in quantum optics (e.g., Ref.~\cite{GK05}).
An operator-valued Schr\"odinger equation is derived and is shown to be equivalent to
Heisenberg equation for field operators,
and the Boson/Fermion statistics arises naturally due to such equivalence.

The operator-valued quantum phase space formula
provides a natural link between the first and second quantization.
We term the process of converting a dynamics into a Hamiltonian dynamics in a phase space as {\em hamiltonization},
for convenience.
A first-quantized system can be further second-quantized through hamiltonization.
This forms a foundation of second quantization.
Furthermore, combining quantization and hamiltonization together,
we can generalize the notion of quantization and break through the framework of the first and second quantization.
Given a dynamics in a certain space, a Hamiltonian dynamics can be constructed,
and the Hamiltonian dynamics can be further quantized to a dynamics in a Hilbert space,
and so on.
As such, a {\em quantization hierarchy} is possible and a series of dynamics on different levels of the
hierarchy can be constructed.

As a more practical application of our quantization method based on operator-valued quantum phase space,
we study the dynamics of quantum many-body system.
A quantum many-body system is usually either described by a first-quantized Hamiltonian
along with density operator theory (DOT) and the notion of (e.g.) entanglement~\cite{HHHH09},
or described by a second-quantized Hamiltonian along with quantum field theory (QFT).
We are concerned with the problem of {\em ab initio} modelling a quantum many-body system.
We show that in principle DOT and QFT provide equivalent models.
That is, a first-quantized system based on DOT
can be second-quantized to a model based on QFT;
and on the other hand,
a second-quantized system based on QFT can be
`de-quantized' to a model based on DOT.
Our method is fundamentally different from those such as the
slave-particle model and Jordan-Wigner transformation~\cite{JW28,AA08}.

We also construct a new model for quantum many-body system,
in which the parameters in the state to be determined scales polynomially
with the number of particles.
The model is {\em eclectic} in that there are both tensor product
and direct sum structures of the Hilbert space,
wherein the interaction determines the tensor product structure,
and the superselection of subsystems determines the direct sum structure.
The dimension of the Hilbert space scales as $(dn)^k$:
$n$ is the number of particles,
$d$ is the local dimension of each particle,
and $k$ is the locality parameter.

This work includes the following parts.
In section~\ref{sec:qphase} we review the basics of phase space and Geometric Quantum Mechanics,
and then we construct the operator-valued quantum phase space.
In section~\ref{sec:qh} we introduce the notion of quantization hierarchy and focus on some of its basic properties.
In section~\ref{sec:manyb} we apply the quantization method to study the relation between DOT and QFT.
An eclectic model for quantum many-body system is constructed.
In section~\ref{sec:dis} we further discuss some implications.

\section{Quantum phase space}
\label{sec:qphase}

\subsection{Preliminary}
\label{subsec:trad}

\subsubsection{Cotangent bundle, symplectic geometry, and Hamiltonian dynamics}
\label{subsubsec:cotbund}

We start from basic notions in manifold theory (e.g., Ref.~\cite{Fra97}).
For a manifold $V$, there exists tangent space $V_a$ for each $a\in V$.
The set of tangent space is called tangent bundle and denoted as $TV$.
The point in $TV$ is specified by a pair of parameters as
$(a,\dot{a})$, $a\in V$, $\dot{a}\in V_a$,
and $\dot{a}:= \frac{da}{dt}$ is the derivative of $a$
with respect to some external parameter $t\in \mathbb{R}^+$.
The $V^*$ ($V_a^*$) denotes the dual space of $V$ ($V_a$),
and the set of $V_a^*$ is called the cotangent bundle $T^*V$.
If $\alpha\in V_a^*$, then $(a,\alpha)\in T^*V$.

The space $T^*V$ is an even-dimensional symplectic manifold equipped with the symplectic form
\begin{equation}\label{}
\omega=e^1\wedge e^{n+1}+e^2\wedge e^{n+2}+\dots+e^n\wedge e^{2n},
\end{equation}
given $\{e^1,\dots,e^{2n}\}$ the basis of $T^*V$. The $\wedge$ denotes wedge product.
A Hamiltonian function $H$ is defined on the symplectic manifold
associated with a vector field $\upsilon_H$ and flow $h_t$ (to be explained below),
and the triplet $(T^*V,\omega,H)$ is called a {\em Hamiltonian dynamics system}.
For the case of time-dependent $H$,
a pre-symplectic form can be defined and the dynamics system takes a similar form.
In this work we assume all Hamiltonians are time-independent for simplicity,
and the generalization is straightforward.

\subsubsection{Hamiltonization and quantization for Classical Mechanics}
\label{subsubsec:classical}

In Classical Mechanics, a particle has a position $q\in\mathbb{R}^3$,
and the Newtonian dynamics describes the trajectory $q(t)$ for time $t\in \mathbb{R}^+$.
The momentum $p$ is associated with the velocity $\dot{q}$ and is defined as
$p=\frac{\partial L}{\partial \dot{q}}$,
with Lagrangian $L(q,\dot{q},t)$.
Denote $V\equiv \mathbb{R}^3$,
then $p\in V_q^*$, and $(q,p)\in T^*V$.
The {\em phase space} is the cotangent bundle $T^*V$ for the particle dynamics specified by parameters $(q,p)$.
The symplectic form is $\omega=dq\wedge dp$.
The Hamiltonian is defined as $H=p\dot{q}-L$, from which, the standard Hamiltonian dynamics follows
\begin{align}\label{}
  \frac{\partial H}{\partial p}=\dot{q}, \;\;
  \frac{\partial H}{\partial q}=-\dot{p}.
\end{align}

The Poisson bracket for two functions $f,g$ takes the form
\begin{equation}\label{}
\{f,g\}=\frac{\partial f}{\partial q}\frac{\partial g}{\partial p}-\frac{\partial g}{\partial q}\frac{\partial f}{\partial p},
\end{equation}
which also satisfies the Jacobi identity
\begin{equation}\label{eq:jac}
\{f,\{g,h\}\}+\{h,\{f,g\}\}+\{g,\{h,f\}\}=0,
\end{equation}
for $f,g,h\in C(T^*V)$, and $C(\cdot)$ denotes the set of functions on the phase space.
The Hamiltonian dynamics can also be expressed as
\begin{align}\label{}
   \{q,H\}=\dot{q}, \;\;
   \{p,H\}=\dot{p}.
\end{align}

The Hamiltonian $H$ induces a Hamiltonian flow $h_t: T^*V\rightarrow T^*V$,
and a vector field
$\upsilon_H :=\frac{\partial H}{\partial p}\frac{\partial }{\partial q}-\frac{\partial H}{\partial q}\frac{\partial }{\partial p}$,
and the Hamiltonian dynamics is equivalent to
\begin{equation}\label{}
  \dot{\eta}=\upsilon_H\eta,
\end{equation}
with the combined column vector $\eta :=(q,p)$.

Quantization is an operation which maps functions on a space
to operators acting on the corresponding Hilbert space $\mathcal{H}$.
For $\mathbb{R}^3$, $\mathcal{H}=L_2(\mathbb{R}^3)$,
the set of square-integrable functions on it
(ignoring the normalization condition).
The dynamical variables $(q,p)$ are mapped to operators
$(\hat{q},\hat{p}:=-i\frac{\partial}{\partial q})$,
which are both hermitian observable,
and the Hamiltonian $H$ is mapped to quantum Hamiltonian operator $\hat{H}$.
The Schr\"odinger equation takes the form
\begin{equation}\label{}
  i|\dot{\psi}\rangle=\hat{H}|\psi\rangle, \; \text{or} \; i\dot{\psi}(x)=\hat{H}\psi(x),
\end{equation}
with $|\psi\rangle=\int dx \psi(x) |x\rangle$, and
$|x\rangle, |\psi\rangle\in \mathcal{H}$, $\psi(x)\in \mathcal{H}^*$,
with $\mathcal{H}^*$ as the dual of $\mathcal{H}$, and we let $\hbar\equiv1$.
In addition, there are issues about unbounded operator, e.g., $\hat{q},\hat{p}$ are not bounded
so that their domain are not the whole Hilbert space,
in which case the Hilbert space is relaxed to allow ``rigged'' Hilbert space~\cite{RM05}.
As this is not the focus of this work, we do not study these technical issues.

\subsection{Geometric Quantum Mechanics}
\label{subsec:gQM}

The research of geometrization of Quantum Mechanics
is mostly motivated by General Relativity and a unification of them.
This formula of quantum theory shows that
quantum system can be treated as a Hamiltonian dynamics system,
while the dynamical variables are different from those in Classical Mechanics.
Although there exist different approaches~\cite{Kibb79,CMP90,CCM07},
which may start from Hilbert space,
projective Hilbert space or the C$^*$-algebra on it,
there are some basic facts of this approach.

For a finite-dimensional Hilbert space $\mathcal{H}$ with orthonormal basis
$\{|i\rangle\}$ and $d=\text{dim}\mathcal{H}< \infty$,
a state $|\psi\rangle =\sum_{i=1}^d \psi_i |i\rangle \in \mathcal{H}$.
Here we do not require the normalization condition,
and also the probability interpretation of quantum state.
For infinite-dimensional $\mathcal{H}$, the sum is changed into an integral.
We focus on the finite-dimensional case.

Hilbert space $\mathcal{H}$
is a K\"ahler manifold with a symplectic form Im$\langle\psi|\phi\rangle$
and a Riemannian form Re$\langle\psi|\phi\rangle$, for $|\psi\rangle, |\phi\rangle \in \mathcal{H}$.
The dynamical variables are $(\psi_i, i\psi_i^*)$, where
the Hamiltonian function is the energy
\begin{equation}\label{}
H=\langle \psi| \hat{H} |\psi\rangle=\sum_{i,j=1}^d\psi^*_i H_{ij}\psi_j.
\end{equation}
We define the ``hamiltonization'' map
\begin{equation}\label{}
\text{Hamiltonization}: \mathcal{H} \rightarrow \Sigma: \hat{H} \mapsto H , |\psi\rangle \mapsto (\psi_i, \zeta_i),
\end{equation}
with $i\psi_i^*\equiv\zeta_i$,
$\Sigma$ denoting the K\"ahler manifold which is actually the Hilbert space $\mathcal{H}$ itself.
The hamiltonization map is a homomorphism instead of an isomorphism
since given a Hamiltonian dynamics different quantum Hamiltonian operators can be deduced.
This is similar with the fact that given a classical dynamics there can be different quantum versions.

The Hamilton's equations on $\Sigma$ are
\begin{equation}\label{eq:gqm}
  \frac{\partial H}{\partial \psi_i}=-\dot{\zeta}_i \; , \; \frac{\partial H}{\partial \zeta_i}= \dot{\psi}_i.
\end{equation}
The symplectic form is $\omega=\sum_{i=1}^d d\psi_i \wedge d \zeta_i$,
and the Hamiltonian dynamics system is $(\Sigma, \omega, H)$.
The Eq.~(\ref{eq:gqm}) can also be expressed via Re$\psi_i$ and Im$\psi_i$.
We can see that a $d$-dimensional quantum particle can be viewed as $d$ coupled classical particles (or excitation).
If the Hilbert space is infinite dimensional, e.g., $L_2(\mathbb{R}^3)$,
the classical dynamical variables become scale field $\psi(x)$ and $i\psi^*(x)$,
and the Hamilton's equations take the form for a scale field.
Note that there is a significant difference between the field here and a scale field in classical wave mechanics,
where the `momentum' is the time-derivative of the field.

Quantum observable corresponds to special kind of vector field, such as the K\"ahlerian function~\cite{CMP90},
and the Poisson bracket between two vector fields is equivalent to the expectation value of commutator
\begin{equation}\label{eq:Pos-comtor}
\{F,G\}=-i\langle\psi|[\hat{F},\hat{G}]|\psi\rangle,
\end{equation}
with observable $\hat{F},\hat{G}$ and the corresponding vector fields $F, G$.

\subsection{Operator-valued quantum phase space}
\label{subsec:ovps}

The Ehrenfest theorem reveals that a classical description of quantum dynamics
can be obtained by taking the expectation value of operators on a quantum state.
From this point of view, the dynamics~(\ref{eq:gqm}) is ``classical'' in the sense that
the Hamiltonian is the expectation values of the quantum operators,
which is consistent with the Ehrenfest theorem.
It turns out we can generalize the form above to the quantum case,
or termed as an operator-valued case wherein the expectation is absent.
That is, we can further quantize the Hamiltonian dynamics on $\Sigma$
to a new quantum dynamics.

The quantization map takes the form
\begin{equation}\label{}
\text{Quantization}: \Sigma\rightarrow L_2(\Sigma):
H\mapsto \mathbb{H}, (\psi_i, \zeta_i) \mapsto (\hat{\psi}_i, \hat{\zeta}_i),
\end{equation}
and the new quantum dynamics is
\begin{equation}\label{}
  i |\dot{\chi}(\psi)\rangle= \mathbb{H}|\chi(\psi)\rangle, \;
  \text{or} \;\; i \dot{\chi} (\psi) = \mathbb{H} \; \chi (\psi),
\end{equation}
with $\chi (\psi) \in L_2(\Sigma)$,
and $|\chi(\psi)\rangle=\int d\psi \chi(\psi) |\psi\rangle$, and
\begin{equation}\label{}
  \mathbb{H}=\hat{\psi}^\dagger \hat{H}\hat{\psi} = \sum_{i,j=1}^d\hat{\psi}^\dagger_i H_{ij} \hat{\psi}_j,
\end{equation}
with $\hat{\psi}=\sum_{i=1}^d \hat{\psi}_i |i\rangle$,
and $\mathbb{H}, \hat{\psi}_i: L_2(\Sigma)\rightarrow L_2(\Sigma)$.
Note $H_{ij}$ could be an operator so the order in $\mathbb{H}$ cannot be changed generally.
We may call $\hat{\psi}$ as field operator.
The form $d\psi$ represents a measure on space $L_2(\Sigma)$.
As is well known, there is no Lebesgue measure on infinite-dimensional Hilbert space; however,
there can be a Borel measure, which is indeed the Wiener measure~\cite{GJ87}.
Also, employing Fock space and particle-number basis or coherent-state basis (studied below)
instead of space $L_2(\Sigma)$,
a measure and integral can be well defined.

The formalism above provides a proper foundation for second quantization.
We can view $\hat{\psi}_i$ as the analog of ladder operator $\hat{a}$
of a position particle and then $\sum_i \hat{\psi}^\dagger_i\hat{\psi}_i$
is the particle (or excitation) number operator.
In order to describe the boson or fermion,
the Fock space $\mathcal{F}$ is often introduced,
which is isomorphic to $L_2(\Sigma)$,
since their dimensions are the same.
As the result, we can study the dynamics in $\mathcal{F}$ instead.
In Fock space, there exists the particle number basis $\{|n\rangle\}$,
and state $|\chi(\psi)\rangle$ can be expanded in this basis as
\begin{equation}\label{}
|\chi\rangle=\sum_{n=0}^\infty \chi_n |n\rangle, \; \; |\chi\rangle \in \mathcal{F}.
\end{equation}

The Hamiltonian $\mathbb{H}$ is a hermitian linear operator.
In literature, there are similar forms called bilinear operator in the approach to second-quantize spin systems~\cite{AA94}.
A bilinear map $B: V\times W \rightarrow X$ is linear with respect to each of the space $V$ and $W$.
However, $\mathbb{H}$ is not bilinear, since $\hat{\psi}$ and
$\hat{\psi}^\dagger$ acts on $ \mathcal{H} \otimes\mathcal{F}$ instead of $\mathcal{F}$,
and as such, $\mathbb{H}$ only acts on the space for $\hat{\psi}_i$ not $\hat{H}$.

Next we study the dynamics of field operators $\hat{\psi}_i$.
We find that the Schr\"odinger equation and Heisenberg equation are equivalent,
and furthermore, the equivalence is closely related to the Boson/Fermion statistics.
Starting from Heisenberg equation of the observable $\hat{\psi}_i$
\begin{equation}\label{}
  i \dot{\hat{\psi}}_i=[\hat{\psi}_i, \mathbb{H}], \; \; \& \; \;
  i \dot{\hat{\psi}}=[\hat{\psi}, \mathbb{H}],
\end{equation}
we find if the operator $\hat{\psi}_i$ is bosonic or fermionic
\begin{subequations}
\label{eq:bosonfermion}
\begin{align}
  \text{Boson:\;} & [\hat{\psi}_i, \hat{\psi}_j]=0, [\hat{\psi}^\dagger_i, \hat{\psi}^\dagger_j]=0 ,
[\hat{\psi}_i, \hat{\psi}^\dagger_j]=\delta_{ij},
\end{align}
\begin{align}
  \text{Fermion:\;} & \{\hat{\psi}_i, \hat{\psi}_j\}=0, \{\hat{\psi}^\dagger_i, \hat{\psi}^\dagger_j\}=0,
\{\hat{\psi}_i, \hat{\psi}^\dagger_j\}=\delta_{ij},
\end{align}
\end{subequations}
then
\begin{equation}\label{eq:schhei}
  i \dot{\hat{\psi}}_i=[\hat{\psi}_i, \mathbb{H}]=\sum_{j=1}^dH_{ij} \hat{\psi}_j, \; \; \& \;  \;
  i \dot{\hat{\psi}}=[\hat{\psi}, \mathbb{H}]=\hat{H} \hat{\psi}.
\end{equation}
The Kronecker delta $\delta_{ij}$ is changed to Dirac delta function $\delta(i-j)$
for infinite-dimensional case.
The equation~(\ref{eq:schhei}) above is the operator-valued version of
$i \dot{\psi}_i=\sum_{j=1}^d H_{ij} \psi_j$ and $i|\dot{\psi}\rangle=\hat{H}|\psi\rangle$.

Conversely, if we assume Eq.~(\ref{eq:schhei}) first
and then we can derive the Boson/Fermion statistics~(\ref{eq:bosonfermion}).
That is, the derivation of statistics does not refer to spin or Special Relativity.
From spin-statistics connection~\cite{MR03},
if $d$ is even, the pseudo-spin is half-integral, and the field is fermionic;
if $d$ is odd, the pseudo-spin is integral, and then the field is bosonic.
Also, the equivalence between Schr\"odinger equation and Heisenberg equation above
is more fundamental than the common one,
which is about the expectation value of observable
in two different pictures.

Furthermore, there is also an operator-valued version of the Hamiltonian dynamics (in Eq.~(\ref{eq:gqm})).
 We find that
\begin{equation}\label{}
  \frac{\partial \mathbb{H}}{\partial \hat{\psi}_i}=-\dot{\hat{\zeta}}_i \; , \; \frac{\partial \mathbb{H}}{\partial \hat{\zeta}_i}= \dot{\hat{\psi}}_i.
\end{equation}
Then the following equation holds
\begin{equation}\label{}
  i \dot{\hat{\psi}}=[\hat{\psi}, \mathbb{H}]=\hat{H} \hat{\psi}=\frac{\partial \mathbb{H}}{\partial \hat{\psi}^\dagger}.
\end{equation}
This equation looks simple, while it may have deep physical foundations.
The first equality is Heisenberg equation, the second one is Schr\"odinger equation,
and the third one is Hamilton equation.
The symplectic form is $\omega=d\hat{\psi}\wedge d\hat{\zeta}=\sum_i d\hat{\psi}_i\wedge d\hat{\zeta}_i$,
and the triplet $(\mathcal{F},\omega,\mathbb{H})$
should be called {\em quantum (or operator-valued) Hamiltonian dynamics system}.

\subsubsection{Observable}

For hermitian operator $\hat{O}$ acting on Hilbert space $\mathcal{H}$,
define the observable acting on $\mathcal{F}$ as
\begin{equation}\label{}
 \mathbb{O} :=\hat{\psi}^\dagger\hat{O}\hat{\psi}.
\end{equation}
A quantum Poisson structure can also be defined as in the case of Classical Mechanics.
For operators $\mathbb{F}$ and $\mathbb{G}$ acting on $\mathcal{F}$,
the quantum Poisson bracket is defined as
\begin{equation}\label{}
\{\mathbb{F},\mathbb{G}\}_Q :=\frac{\partial\mathbb{F}}{\partial \hat{\psi}}\frac{\partial \mathbb{G}}{\partial \hat{\zeta}}-\frac{\partial \mathbb{G}}{\partial \hat{\psi}}\frac{\partial \mathbb{F}}{\partial \hat{\zeta}}.
\end{equation}
It is straightforward to check that
$[\mathbb{O},\mathbb{H}]=i\{\mathbb{O},\mathbb{H}\}_Q$, and more generally,
\begin{equation}\label{}
  [\mathbb{F},\mathbb{G}]=i\{\mathbb{F},\mathbb{G}\}_Q.
\end{equation}
We see that the commutator $[,]$ is equivalent to quantum Poisson bracket $\{,\}_Q$,
which means that the commutator plays the roles of quantum Poisson bracket.
This is different from the common view that
the expectation value of commutator corresponds to the classical Poisson bracket,
also see Eq.~(\ref{eq:Pos-comtor}).

The quantum Poisson bracket also satisfies the Jacobi identity
\begin{equation}\label{}
\{\mathbb{F},\{\mathbb{G},\mathbb{E}\}\}_Q+\{\mathbb{E},\{\mathbb{F},\mathbb{G}\}\}_Q+\{\mathbb{G},\{\mathbb{E},\mathbb{F}\}\}_Q=0,
\end{equation}
which is equivalent to the Jacobi identity of the corresponding operators acting on Hilbert space; i.e.
$[F,[G,E]]+[E,[F,G]]+[G,[E,F]]=0$.

Next we consider observable dynamics in Heisenberg picture.
Define time-dependent observable $\hat{O}_t:=e^{i\hat{H}t}\hat{O}e^{-i\hat{H}t}$, and then
$\mathbb{O} :=\hat{\psi}^\dagger_t\hat{O}\hat{\psi}_t=\hat{\psi}^\dagger_0\hat{O}_t\hat{\psi}_0$.
We find
\begin{equation}\label{eq:mix2}
  i\dot{\mathbb{O}}=[\mathbb{O},\mathbb{H}]=\hat{\psi}_t^\dagger[\hat{O},\hat{H}]\hat{\psi}_t
  =\hat{\psi}_0^\dagger[\hat{O}_t,\hat{H}]\hat{\psi}_0,
\end{equation}
from which we derive
\begin{equation}\label{}
 \dot{\hat{O}}_t=-i[\hat{O}_t,\hat{H}],
\end{equation}
which is the usual Heisenberg equation, assuming $\dot{\hat{O}}=0$.
Also, the Hamiltonian-Poissonian dynamics of field operator can be summarized as
\begin{subequations}\label{}
\begin{equation}
\dot{\hat{\psi}} =  \frac{\partial \mathbb{H}}{\partial \hat{\zeta}}=-i \hat{H} \hat{\psi} = -i[\hat{\psi},\mathbb{H}]=\{\hat{\psi},\mathbb{H}\}_Q,
\end{equation}
\begin{equation}\label{}
  \dot{\hat{\zeta}} = -\frac{\partial \mathbb{H}}{\partial \hat{\psi}}=i \hat{\zeta}\hat{H}= -i[\hat{\zeta},\mathbb{H}]=\{\hat{\zeta},\mathbb{H}\}_Q.
\end{equation}
\end{subequations}

\subsubsection{State}

Next we derive the dynamics of state in Schr\"odinger picture.
A quantum state is generally described by a density operator $\hat{\rho}$.
As $\hat{\rho}=\hat{\rho}^\dagger$, a direct application of the Hamiltonization process does not work.
Instead, we second-quantize $\hat{\rho}$
as $\hat{\rho}\mapsto \hat{\varrho}\equiv\hat{\psi}\hat{\psi}^\dagger$ for pure state case,
and $\hat{\rho}\mapsto \hat{\varrho}\equiv\sum_\mu p_\mu \hat{\psi}_\mu\hat{\psi}^\dagger_\mu$ for mixed state case,
with eigen-decomposition
$\rho=\sum_\mu p_\mu |\psi_\mu\rangle\langle \psi_\mu|$, $\sum_\mu p_\mu =1$, $p_\mu\in (0,1)$.
Although density operator is hermitian, it cannot be simply treated as a positive observable.
That is, there are foundational differences between density operator and observable,
which, e.g., is manifested in the C$^*$-algebra approach~\cite{Segal47}.
There is a ``minus-sign'' difference between their dynamical equation, which indicates that
the quantum Poisson bracket needs to be modified as
\begin{equation}\label{}
\{\mathbb{F},\mathbb{G}\}_{\bar{Q}} :=\frac{\partial\mathbb{G}}{\partial \hat{\psi}}\frac{\partial \mathbb{F}}{\partial \hat{\zeta}}-\frac{\partial \mathbb{F}}{\partial \hat{\psi}}\frac{\partial \mathbb{G}}{\partial \hat{\zeta}},
\end{equation}
and the Jacobi identity still holds.
Notice that $\hat{\psi}$ and $\hat{\psi}^\dagger$ are normal ordered in observable,
while antinormal ordered in state.

In the following, we study the pure state case first and then generalize it to mixed state case.
For pure state case, we find
\begin{equation}\label{eq:mix}
  i \dot{\hat{\varrho}}=[\hat{H}, \hat{\varrho}]=i\{\hat{\varrho},\mathbb{H}\}_{\bar{Q}}.
\end{equation}
The first equality is the operator-valued version of von Neumann equation,
and the second equality is the quantum version of classical Liouvillian equation,
and $[\hat{H}, \hat{\varrho}]=i\{\hat{\varrho},\mathbb{H}\}_{\bar{Q}}$
shows the equivalence between commutator and quantum Poisson bracket.

For mixed state case, define
\begin{align}\label{}
  \{\mathbb{F},\mathbb{G}\}_{\bar{Q}} &:=\sum_\mu \frac{1}{p_\mu}
  \left[\frac{\partial \mathbb{G}}{\partial\hat{\psi}_\mu}
  \frac{\partial \mathbb{F}}{\partial\hat{\zeta}_\mu} -\frac{\partial \mathbb{F}}{\partial\hat{\psi}_\mu}
  \frac{\partial \mathbb{G}}{\partial\hat{\zeta}_\mu}\right], \\
  \mathbb{H} &:=\sum_\mu p_\mu \hat{\psi}_\mu^\dagger\hat{H}\hat{\psi}_\mu.
\end{align}
It is direct to check that Equation~(\ref{eq:mix}) above still holds.
That is, we can naturally generalize the quantum symplectic space to the mixed state case.

\subsubsection{Nonunitary evolution}
\label{sec:open}

More generally, quantum dynamics can be nonunitary, such as completely positive semigroup~\cite{Bre03}.
The standard Lindblad equation~\cite{Lind76} for observable dynamics takes the form
\begin{equation}\label{}
  \dot{\hat{O}}=-i[\hat{O},\hat{H}]+\sum_\alpha \gamma_\alpha \left( \hat{L}_\alpha^\dagger \hat{O}\hat{L}_\alpha-\frac{1}{2}\{\hat{L}_\alpha^\dagger\hat{L}_\alpha,\hat{O}\} \right),
\end{equation}
for a set of Lindblad operators $\hat{L}_\alpha$ and decay coefficients $\gamma_\alpha$.
We find the second-quantized form is
\begin{equation}\label{}
  \dot{\mathbb{O}}=-i[\mathbb{O},\mathbb{H}]+\sum_\alpha \gamma_\alpha \hat{\psi}^\dagger\left( \hat{L}_\alpha^\dagger \hat{O}\hat{L}_\alpha-\frac{1}{2}\{\hat{L}_\alpha^\dagger\hat{L}_\alpha,\hat{O}\} \right)\hat{\psi}.
\end{equation}

Similarly,
we obtain the second-quantized form for the dynamics of quantum state
\begin{equation}\label{}
  \dot{\hat{\varrho}}=i[\hat{\varrho},\hat{H}]+\sum_\alpha \gamma_\alpha \left( \hat{L}_\alpha \hat{\varrho}\hat{L}_\alpha^\dagger-\frac{1}{2}\{\hat{L}_\alpha^\dagger\hat{L}_\alpha,\hat{\varrho}\} \right).
\end{equation}

For completely positive map $\mathcal{E}$~\cite{NC00}, quantum state evolves as
$\mathcal{E}(\hat{\rho})=\sum_\alpha K_\alpha\hat{\rho} K_\alpha^\dagger$.
The second-quantized form is $\mathcal{E}(\hat{\varrho})=\sum_\alpha K_\alpha\hat{\varrho} K_\alpha^\dagger$.
In Heisenberg picture, the action of the map on observable is
$\mathcal{E}(\hat{O})=\sum_\alpha K_\alpha^\dagger\hat{O} K_\alpha$.
The second-quantized form is $\mathcal{E}(\mathbb{O})=\sum_\alpha K_\alpha^\dagger \mathbb{O} K_\alpha$.
Note here the map $\mathcal{E}$ acts on $\mathcal{B}(\mathcal{H})$,
while $\hat{\varrho}$ and $\mathbb{O}$ act on $\mathcal{B}(\mathcal{H})\otimes \mathcal{F} $.

Despite the above general formula, the most desirable equation is the dynamics for field operator $\hat{\psi}$.
However, it is difficult (or impossible) to derive the field operator dynamics
since the dynamics of $\hat{\varrho}$ couples $\hat{\psi}$ and $\hat{\psi}^\dagger$ together.
As a result, a more suitable formula to start with is the stochastic Schor\"odinger equation~\cite{Bre03},
in which the dynamical variable is quantum state $|\psi\rangle$ instead of $\hat{\rho}$.
The Lindblad master equation can be derived from stochastic Schor\"odinger equation
by taking ensemble average.
The second-quantized form of stochastic Schor\"odinger equation
can be easily obtained by substituting the quantum state vector by the field operator.

\section{Quantization Hierarchy}
\label{sec:qh}

In the last section we have developed the operator-valued quantum phase space formalism, from which we have
derived the Boson/Fermion statistics and provided a link between first and second quantization.
In this section, we explore more implications of our quantization method.

Based on hamiltonization,
we can see that dynamics in $\mathcal{F}$ also forms a classical Hamiltonian system,
with the Hamiltonian function $\langle\chi|\mathbb{H}|\chi\rangle$ and dynamical variable $\chi(\psi)$,
which is a scale field.
The new phase space Hamiltonian dynamics can be further quantized and then again be hamiltonized.
Based on this procedure, there exists the quantization hierarchy:
\begin{equation}\label{}
  \Sigma_0 \rightarrow \mathcal{H}_1 \rightarrow \Sigma_1 \rightarrow \mathcal{H}_2 \rightarrow \Sigma_2 \rightarrow \mathcal{H}_3 \rightarrow \cdots
\end{equation}
when it starts from a classical system in space $\Sigma_0$, or
\begin{equation}\label{}
   \mathcal{H}_0  \rightarrow \Sigma_0 \rightarrow \mathcal{H}_1 \rightarrow \Sigma_1 \rightarrow \mathcal{H}_2 \rightarrow \Sigma_2  \rightarrow \cdots
\end{equation}
when it starts from a quantum system in Hilbert space $\mathcal{H}_0$.

A qualitative understanding of hierarchy is as follows.
Consider the hierarchy $\Sigma_0 \rightarrow \mathcal{H}_1 \rightarrow \Sigma_1 \rightarrow \cdots$.
If a quantum dynamics is given with some Hilbert space $\mathcal{H}_i$,
then it can be converted into a Hamiltonian system in space $\Sigma_i$ by hamiltonization,
or it can be reduced to the classical dynamics in space $\Sigma_{i-1}$ by taking expectation, based on Ehrenfest theorem.
The dynamical variables in space $\Sigma_{i-1}$ and $\Sigma_i$ are different.
If a classical dynamics is given in space $\Sigma_i$,
then we need to justify which level it is in the hierarchy.
The dynamics in $\Sigma_i$ can be quantized to a higher level in space $\mathcal{H}_{i+1}$,
or it could be used to recover the quantum dynamics in space $\mathcal{H}_i$,
and both of the processes are not unique.
The dynamics in $\Sigma_i$ is equivalent to the dynamics in $\mathcal{H}_{i+1}$ by taking expectation in $\mathcal{H}_{i+1}$,
and the dynamics in $\mathcal{H}_{i+1}$ is equivalent to the dynamics in  $\Sigma_{i+1}$ by hamiltonization,
so the dynamics in all the levels are equivalent to each other.
The analysis is similar for the other hierarchy $\mathcal{H}_0 \rightarrow \Sigma_0 \rightarrow \mathcal{H}_1  \rightarrow \cdots$.

The equivalence between the first and second quantization is a special case of the property above.
For instance, if there is a ``first-level'' quantum dynamics $i|\dot{\psi}\rangle=\hat{H}_1|\psi\rangle$ with Hamiltonian $\hat{H}_1$,
and in ``second-level'' we have $i|\dot{\chi}\rangle=\hat{H}_2|\chi\rangle$
with $\hat{H}_2=\sum_{ij}\hat{\psi}^\dagger_i H_{ij}\hat{\psi}_j$.
Then for the energy, there exists a second-level state $|\chi\rangle$ such that $\langle\psi|\hat{H}_1|\psi\rangle=\langle\chi|\hat{H}_2|\chi\rangle$.
It is clear that their numerical ranges are the same.
However, there is a crucial property that
the spectrum of $\hat{H}_1$ is not the same with $\hat{H}_2$, and
the ground state of $\hat{H}_1$ may not correspond to the ground state of $\hat{H}_2$.
This can be easily generalized to mixed state case, e.g.,
there exist states $\rho$ and $\sigma$ such that $\text{tr}(\rho \hat{H}_1)=\text{tr}(\sigma \hat{H}_2)$.

In the hierarchy, the dimension of Hilbert space will become infinite from a certain level,
although it may start from a finite-dimensional Hilbert space.
The Hamiltonian dynamics will become that for a scale field instead of a finite collection of particles (excitations).
Indeed, one crucial foundation for the existence of such hierarchy is that
infinite-dimensional Hilbert spaces are isomorphic with each other.

The analysis of quantization hierarchy would become complicated when interaction exists,
since systems at different levels of their quantization hierarchies can interact with each other.
For instance, a system in first-quantized form can interact with another system in second-quantized form,
while only the latter will respect the Boson/Fermion statistics.
A qubit (spin-$1/2$ system) interacting with a harmonic oscillator is one example of this type.
Furthermore, for a fixed interaction,
an interacting subsystem cannot be changed from one level of its hierarchy to another level,
although different levels of this subsystem are equivalent.
The reason is that the interaction among the subsystems has already ``select'' a specific level of one subsystem in its hierarchy.
If we change the quantization level of one subsystem, the interaction needs to be modified.
To decide the formula of interaction is a nontrivial problem,
and particularly, it is a deep question whether
interaction induces a ``selection'' of a special level on the hierarchy of a system,
or a system ``spontaneously'' selects a special level and then interacts with other systems.

\subsection{Examples}
\label{subsec:example}

Next, we analyze some examples to justify the method of quantization hierarchy.

\begin{example}
Harmonic oscillator in $\mathbb{R}$ or $\mathbb{C}$.

\emph{Consider a single harmonic oscillator with Hamiltonian $H=x^2+p^2$ (ignoring other coefficients).
Usually, it is quantized as $\hat{H}=\hat{x}^2+\hat{p}^2$ with observable $\hat{x}$ and $\hat{p}$.
Given  $\hat{x}$ and $\hat{p}$, the ladder operators $\hat{a}=\hat{x}+i\hat{p}$ and $\hat{a}^\dagger$ are introduced
and then $\hat{H}=(\hat{a}^\dagger \hat{a}+1/2)$.
A disturbing fact is that the quantum Hamiltonian takes a ``second quantized'' form,
yet it should be only a ``first quantized'' form since it comes from a classical system.
To resolve this puzzle, using the program of quantization hierarchy,
there should be a lower-level quantum system associated with the classical system.
We find, indeed, it is reasonable to assume there is a ``pseudo'' lower-level quantum system whose expectation
generates the classical dynamics.
Also, in this way we can get rid of the zero-point energy.
The hierarchy is shown below.}

\emph{$\mathbb{C}$ ($\mathcal{H}_0$):
Let $a=x+ip \in \mathbb{C}$ with $x\in \mathbb{R}$ and $p\in T_x\mathbb{R}$.
The Hamiltonian is $\hat{H}_0=\hbar \omega$.
The parameter $a$ plays the roles of wavefunction.\\
$\rightarrow \Sigma_0$: $H_0=\hbar\omega a^* a=x^2+p^2$.
Hamiltonian dynamics $\frac{\partial H_0}{\partial x}=-\dot{p}$, $\frac{\partial H_0}{\partial p}=\dot{x}$. \\
$\rightarrow \mathcal{H}_1$: $\hat{H}_1=\hbar\omega \hat{a}^\dagger \hat{a}=\hat{x}^2+\hat{p}^2$,
Quantum dynamics $i\dot{\phi}(x)=\hat{H}_1\phi(x)$ or $i|\dot{\phi}\rangle=\hat{H}_1|\phi\rangle$.
Observable dynamics, e.g., $i\dot{\hat{x}}=[\hat{x},\hat{H}_1]=\hbar\omega\hat{x}=i\frac{\partial \hat{H}_1}{\partial \hat{p}}$.\\
$\rightarrow \Sigma_1$: $H_1=\int dx \phi^*(x) \hat{H}_1\phi(x)$.
Hamiltonian dynamics $\frac{\partial H_1}{\partial \phi(x)}=-i\dot{\phi}^*(x)$,
$\frac{\partial H_1}{\partial \phi^*(x)}=i\dot{\phi}(x)$. \\
$\rightarrow \mathcal{H}_2$: $\hat{H}_2=\int dx \hat{\phi}^\dagger(x) \hat{H}_1\hat{\phi}(x)$, \; \dots \\
$\rightarrow \Sigma_2$ \dots}

\emph{The Hamiltonian $\hat{H}_2$ describes the dynamics of a quantum scale field,
which is equivalent to infinite number of particles,
while the starting Hamiltonian $\hat{H}_1$ is for a single particle.
Furthermore, the single-particle Hamiltonian $\hat{H}_1$ describes any number of phonon,
while the Hamiltonian $\hat{H}_0$ is for a single phonon.
There is no interaction among the phonons in $\hat{H}_1$,
and there is no interaction among the particles in $\hat{H}_2$, neither.
If there does exist interaction in $\hat{H}_2$,
and then this means $\hat{H}_1$ has to be a many-body Hamiltonian.}
\hfill $\blacksquare$
\end{example}

\begin{example}
Position particle in a potential.

\emph{Consider a single particle with classical Hamiltonian $H=p^2+V(x)$,
such as an object in gravitational potential with $V(x)\propto -\frac{1}{x}$.
This case is much harder than the case of harmonic oscillator,
in which the Hamiltonian $\hat{H}_0$ is proportional to identity.
Assume the transformation between $(x,p)$ and $(a,a^*)$
still holds, let $H_0= a^*\hat{H}_0 a$.
Observe that $V(x)$ can be converted to a polynomial of $a^*$ and $a$,
then $\hat{H}_0$ can be constructed as a many-body interaction.}

\emph{Compared to quantization towards lower levels,
the quantization to higher-level ones is much easier,
which is also the commonly used method in standard quantum theory.}

\emph{$\mathbb{C}$ ($\mathcal{H}_0$): Let $a=x+ip \in \mathbb{C}$ with $x\in \mathbb{R}$ and $p\in T_x\mathbb{R}$.
The Hamiltonian is $\hat{H}_0$.\\
$\rightarrow \Sigma_0$: $H_0=a^*\hat{H}_0 a=p^2+V(x)$.
Dynamics $\frac{\partial H_0}{\partial x}=-\dot{p}$, $\frac{\partial H_0}{\partial p}=\dot{x}$. \\
$\rightarrow \mathcal{H}_1$: $\hat{H}_1=\hat{p}^2+\hat{V}$, Dynamics $i\dot{\phi}(x)=\hat{H}_1\phi(x)$ or $i|\dot{\phi}\rangle=\hat{H}_1|\phi\rangle$.
Observable dynamics e.g. $i\dot{\hat{x}}=[\hat{x},\hat{H}_1]=\hat{H}_0\hat{x}=i\frac{\partial \hat{H}_1}{\partial \hat{p}}$.\\
$\rightarrow \Sigma_1$: $H_1=\int dx \phi^*(x) \hat{H}_1\phi(x)$.
Dynamics $\frac{\partial H_1}{\partial \phi(x)}=-i\dot{\phi}^*(x)$,
$\frac{\partial H_1}{\partial \phi^*(x)}=i\dot{\phi}(x)$. \\
$\rightarrow \mathcal{H}_2$: $\hat{H}_2=\int dx \hat{\phi}^\dagger(x) \hat{H}_1\hat{\phi}(x)$,\dots \\
$\rightarrow \Sigma_2$ \dots}
\hfill $\blacksquare$
\end{example}

Next we consider a hierarchy with a quantum system as starting level.
Let the particle be a qubit, e.g., a spin-$1/2$ system.
This case is straightforward since the hierarchy starts from the hamiltonization map
and then proceeds in the same direction to higher levels.

\begin{example}
Qubit in $\mathbb{C}^2$.

\emph{$\mathbb{C}^2$ ($\mathcal{H}_0$):
Let $|\psi\rangle \in \mathbb{C}^2$, the quantum dynamics is $i|\dot{\psi}\rangle=\hat{H}_0|\psi\rangle$.\\
$\rightarrow \Sigma_0$: $H_0=\sum_{ij} \psi_i^* H_{ij} \psi_j$. Dynamics $\frac{\partial H_0}{\partial \psi_i}=-i\psi_i^*$, $\frac{\partial H_0}{\partial \psi^*_i}=i\dot{\psi}_i$. \\
$\rightarrow \mathcal{H}_1$: $\hat{H}_1=\sum_{ij}\hat{\psi}_i^\dagger H_{ij}\hat{\psi}_j$. Dynamics $i|\dot{\chi}\rangle=\hat{H}_1|\chi\rangle$ or $i\dot{\chi}(\psi)=\hat{H}_1\chi(\psi)$. \\
$\rightarrow \Sigma_1$: $H_1=\int d\psi \chi^*(\psi) \hat{H}_1\chi(\psi)$. Dynamics $\frac{\partial H_1}{\partial \chi(\psi)}=-i\dot{\chi}^*(\psi)$, $\frac{\partial H_1}{\partial \chi^*(\psi)}=i\dot{\chi}(\psi)$. \\
$\rightarrow \mathcal{H}_2$: $\hat{H}_2=\int d\psi \hat{\chi}(\psi)^\dagger \hat{H}_1\hat{\chi}(\psi)$,\dots \\
$\rightarrow \Sigma_2$ \dots}
\hfill $\blacksquare$
\end{example}

In practice, the lower levels of the hierarchy are well studied and being applied in different fields.
The map $\mathcal{H}_1 \rightarrow \Sigma_1$ (hamiltonization) and $\Sigma_1\rightarrow \mathcal{H}_2$ (second quantization)
are well understood.
The map $\Sigma_0\rightarrow \mathcal{H}_1$ is the first quantization, and
the inverse of it, $\Sigma_0\dashleftarrow \mathcal{H}_1$, is the Ehrenfest theorem,
and one notable example is the wave packet dynamics widely used in quantum chemistry~\cite{GS95}.
The Bohmian mechanics also falls into this category, e.g. in double-slit interference experiment
the trajectory of each particle (described by so-called guiding equation) is the expectation of the dynamics of position operator~\cite{Bohm52,Bohm522}.
For inner degree of freedom, representing a spin operator by a vector is also a de-quantization.
For instance, a unitary evolution of a spin is translated to an orthogonal rotation of a vector,
and a nonunitary evolution often becomes an affine map~\cite{FA99}.

The term ``first'' and ``second'' quantization is relative since they depend on the quantization histories
in the hierarchy.
The higher-level quantization is not used so far, although there are some works in literature on ``third quantization'' with different contents~\cite{AS89,MM92,IR06,TP08}.
Our toy model of harmonic oscillator can be viewed as a primary example of higher-level quantized system.
Due to the hierarchy, we can see how an infinite-dimensional dynamics ``emerges'' from a finite-dimensional dynamics.

\section{Quantum many-body system}
\label{sec:manyb}

In this section, we switch to study a more practical problem based on our quantization method.
A new model to describe quantum many-body system is constructed.
Our study is motivated by several observations.
First, the Hilbert space with a tensor-product structure is too large for a local interaction~\cite{PQS+11},
which can only generate an exponentially small volume of the whole Hilbert space within an efficient time.
Second, there is no principle to specify when a system can be modeled by a first or second quantized Hamiltonian,
and why a first quantized system can be mapped to another second quantized model, and vice versa.

\subsection{Equivalence between DOT and QFT}
\label{sec:dotqft}

For a system in an infinite-dimensional Hilbert space described by DOT,
e.g. a collection of interacting particles in $\mathbb{R}^3$,
the way to convert it to a QFT system is well understood.
However, there is no standard way for a finite-dimensional system such as a spin system.
For this case, the most general consideration is a $n$-qudit system.
We show that given a system described by DOT, it can be converted to a system described by QFT.
Also, a system described by QFT can be converted to one by DOT.
The DOT and QFT models are equivalent with respect to observation expectation value.

Consider a $n$-qudit $k$-local Hamiltonian
\begin{equation}\label{}
  \hat{H}=\sum_{l=1}^m \hat{H}_l,
\end{equation}
each $\hat{H}_l$ acting on at most $k$ subsystems, and
$\hat{H}:\mathcal{H}\rightarrow \mathcal{H}$, $\dim(\mathcal{H})=d^n$, $m=O(n^k)$.
The $k$-locality assumption of the interaction is necessary
since highly nonlocal interaction is not likely to exist.
For the nonlocal case we cannot reduce the dimension of the system (studied below),
so we focus on the case of local interaction.
In the following, we loosely use the sum symbols without specifying the upper and lower bound for each case for the ease of notation.

If there are $m_{k'}$ $k'$-local terms (with $\sum_{k'} m_{k'}=m$), then
\begin{align}\label{}\nonumber
  \hat{H}&=\sum_{l_1}^{m_1} \hat{H}_{l_1}^{[1]}+ \sum_{l_2}^{m_2} \hat{H}_{l_2}^{[2]}+\cdots +\sum_{l_k}^{m_k} \hat{H}_{l_k}^{[k]} \\
  &\equiv \hat{H}^{[1]}+\hat{H}^{[2]}+\cdots +\hat{H}^{[k]}.
\end{align}

We consider the energy of the system on state $|\psi\rangle\in \mathcal{H}$, which takes the form
\begin{equation}\label{eq:state}
|\psi\rangle=\sum_{i_1,\cdots, i_n=0}^{d-1} \psi_{i_1,\dots,i_n}|i_1,i_2,\cdots, i_n\rangle
\end{equation}
for $\sum_{i_1,\cdots, i_n=0}^{d-1} |\psi_{i_1,\dots,i_n}|^2=1$.
For the 1-local term $\hat{H}_{l_1}^{[1]}$,
the energy takes the form
\begin{equation}\label{}
  H_{l_1}^{[1]}=\langle\psi|\hat{H}_{l_1}^{[1]}|\psi\rangle
  =\sum_{i_1 j_1} \langle\psi_{l_1;i_1}| \psi_{l_1;j_1}\rangle H_{l_1;i_1j_1}^{[1]},
\end{equation}
for
\begin{equation}\label{}
  |\psi_{l_1;i_1}\rangle:= \sum_{i_2, \cdots,i_n} \psi_{i_1,\dots,i_n}|i_2,\cdots,i_n\rangle,
\end{equation}
which satisfies $\sum_{i_1} \langle \psi_{l_1;i_1}| \psi_{l_1;i_1}\rangle=1$.

The vectors $\{|\psi_{l_1;i_1}\rangle\}$ can be mapped to numbers $\{\psi_{l_1;i_1}\}$,
not uniquely, such that $\langle \psi_{l_1;i_1}|\psi_{l_1;j_1}\rangle=\psi_{l_1;i_1}^*\psi_{l_1;j_1}$.
Then one can define 1-local state
\begin{equation}\label{}
   |\psi_{l_1}^{[1]}\rangle:= \sum_{i_1} \psi_{l_1;i_1}|i_1\rangle,
\end{equation}
and the energy becomes
\begin{equation}\label{eq:1energy}
  H_{l_1}^{[1]} =\sum_{i_1 j_1} \psi_{l_1;i_1}^*  H_{l_1;i_1j_1}^{[1]} \psi_{l_1;j_1} =\langle\psi_{l_1}^{[1]}|\hat{H}_{l_1}^{[1]}|\psi_{l_1}^{[1]}\rangle.
\end{equation}

Next we consider the 2-local terms.
The energy takes the form
\begin{equation}\label{}
  H_{l_2}^{[2]}=\sum_{i_1,i_2,j_1,j_2} \langle\psi_{l_2;i_1i_2}|\psi_{l_2;j_1j_2}\rangle H_{l_2;i_1i_2j_1j_2}^{[2]},
\end{equation}
for
\begin{equation}\label{}
  |\psi_{l_1;i_1i_2}\rangle:= \sum_{i_3, \cdots,i_n} \psi_{i_1,\dots,i_n}|i_3,\cdots,i_n\rangle,
\end{equation}
which satisfies $\sum_{i_1i_2} \langle \psi_{l_2;i_1i_2}| \psi_{l_2;i_1i_2}\rangle=1$.
Here the two interacting particles are considered as a whole.
As the case of 1-local term,
the vectors $\{|\psi_{l_2;i_1i_2}\rangle\}$ can also be mapped to numbers $\{\psi_{l_2;i_1i_2}\}$
such that $\langle \psi_{l_2;i_1i_2}|\psi_{l_2;j_1j_2}\rangle=\psi_{l_2;i_1i_2}^*\psi_{l_2;j_1j_2}$.
Then one can define 2-local state
\begin{equation}\label{}
   |\psi_{l_2}^{[2]}\rangle:= \sum_{i_1i_2} \psi_{l_2;i_1i_2}|i_1i_2\rangle,
\end{equation}
and the energy becomes
\begin{equation}\label{eq:2energy}
  H_{l_2}^{[2]}=\sum_{i_1,i_2,j_1,j_2} \psi_{l_2;i_1i_2}^*H_{l_2;i_1i_2j_1j_2}^{[2]}\psi_{l_2;j_1j_2}
  =\langle\psi_{l_2}^{[2]}|\hat{H}_{l_2}^{[2]}|\psi_{l_2}^{[2]}\rangle.
\end{equation}

The 1-local term can be viewed as a special case of 2-local term,
so we can start from the 2-local terms and then reduce to the 1-local terms.
The Hamiltonian $\hat{H}_{l_1}^{[1]}$ can be extended as $\hat{H}_{l_1}^{[1]}\otimes \openone$.
In Eq.~(\ref{eq:2energy}), if $\hat{H}_{l_2}^{[2]}=\hat{H}_{l_1}^{[1]}\otimes \openone$,
and from the relation $\sum_{i_2} \psi_{l_2;i_1i_2}^* \psi_{l_2;j_1i_2}=\psi_{l_1;i_1}^* \psi_{l_1;j_1}$,
the energy $H_{l_2}^{[2]}$ reduces to $H_{l_1}^{[1]}$ in Eq.~(\ref{eq:1energy}).
Generally, from $\hat{H}_{l_k}^{[k]}=\hat{H}_{l_{k-1}}^{[k-1]}\otimes \openone$,
and
\begin{equation}\label{}
\sum_{i_k} \psi_{l_k;i_1i_2\cdots i_k }^* \psi_{l_k;j_1j_2\cdots j_{k-1}i_k}
=\psi_{l_{k-1};i_1i_2\cdots i_{k-1}}^* \psi_{l_{k-1};j_1j_2\cdots j_{k-1}},
\end{equation}
a $(k-1)$-local term can be embedded into a $k$-local term.
Furthermore, all local terms can be viewed as special cases of $k$-local terms.

Then we consider how to introduce field operator.
There are two ways in standard approaches:
one is to use vacuum state $|0\rangle$ acting on which the field operator $\psi^\dagger_i$ can generate a state $|i\rangle$,
the other is to use transformation to convert first-quantized operators
to field operators.
Our method is different, which is based on quantization of Hamiltonian dynamics.

From Hamiltonization, the energy $H$ generates a Hamiltonian dynamics,
and the parameters $\psi_{l_{k'};j_1j_2\cdots j_{k'}}$ and conjugates are the dynamical variables.
The Hamiltonian dynamics can be further quantized to yield the second-quantized form
\begin{align}\label{eq:secH}
  \mathbb{H}&=\sum_{l_1}^{m_1} \sum_{i_1,j_1}^d \hat{\psi}_{l_1;i_1}^\dagger H_{l_1;i_1 j_1}^{[1]}\hat{\psi}_{l_1;j_1}
  +\sum_{l_2}^{m_2}\sum_{i_1,i_2,j_1,j_2}^{d}\hat{\psi}_{l_2;i_1i_2}^\dagger H^{[2]}_{l_2;i_1i_2 j_1j_2}\hat{\psi}_{l_2;j_1j_2}
  +\cdots  \\ \nonumber
  &+\sum_{l_k}^{m_k}\sum_{i_1\cdots i_k,j_1\cdots j_k}^{d}\hat{\psi}_{l_k;i_1i_2\cdots i_k}^\dagger H^{[k]}_{l_k;i_1\cdots i_k j_1\cdots j_k}\hat{\psi}_{l_k;j_1j_2\cdots j_k},
\end{align}
where $\hat{\psi}_{l_k;j_1j_2\cdots j_k}$ is a $k$-body field operator.
The Hamiltonian $\mathbb{H}$ acts on Fock space $\mathcal{F}$.
Till now we have converted the DOT description of the system into a QFT form.
The DOT and QFT dynamics are equivalent in that
\begin{equation}\label{eq:equiv}
\langle\psi|\hat{H}|\psi\rangle=\langle\chi|\mathbb{H}|\chi\rangle,
\end{equation}
for some quantum state $|\chi\rangle\in\mathcal{F}$.
There also exist such equivalence for other observables.
Note the field operators are not independent with each other.

The formula (\ref{eq:secH}) contains $k$-body field operators,
which can be a polynomial of 1-body field operators $\hat{\psi}_{l_1;i_1}$.
If the field operator is separable (i.e. the quantum state is a product state),
$\hat{\psi}_{l_k;j_1j_2\cdots j_k}=\hat{\psi}_{l_k;j_1}\cdots\hat{\psi}_{l_k;j_k}$,
the Hamiltonian reduces to
\begin{align}\label{}
  \mathbb{H}_\text{sep}&=\sum_{l_1}^{m_1} \sum_{i_1,j_1}^d \hat{\psi}_{l_1;i_1}^\dagger H_{l_1;i_1 j_1}^{[1]}\hat{\psi}_{l_1;j_1}
  +\sum_{l_2}^{m_2}\sum_{i_1,i_2,j_1,j_2}^{d}\hat{\psi}_{l_2;i_2}^\dagger\hat{\psi}_{l_2;i_1}^\dagger H^{[2]}_{l_2;i_1i_2 j_1j_2}\hat{\psi}_{l_2;j_1}\hat{\psi}_{l_2;j_2}
  +\cdots  \\ \nonumber
  &+\sum_{l_k}^{m_k}\sum_{i_1\cdots i_k,j_1\cdots j_k}^{d}\hat{\psi}_{l_k;i_k}^\dagger\cdots\hat{\psi}_{l_k;i_1}^\dagger H^{[k]}_{l_k;i_1\cdots i_k j_1\cdots j_k} \hat{\psi}_{l_k;j_1}\cdots\hat{\psi}_{l_k;j_k}.
\end{align}
It satisfies $\langle\psi|\hat{H}|\psi\rangle=\langle\psi|\mathbb{H}_\text{sep}|\psi\rangle$,
which means one can treat $\hat{H}$ and $\mathbb{H}_\text{sep}$ equally,
as in standard approach of QFT.
However, in general $\hat{H}$ and $\mathbb{H}$ are not the same, and the equivalence between them
is expressed in Eq.~(\ref{eq:equiv}) with respect to expectation value.

\subsection{Eclectic model of many-body system}
\label{sec:eclectic}

Now we have converted a DOT system to a QFT system (Eq.~(\ref{eq:secH}));
furthermore, it turns out the QFT system can be converted into another DOT system.
The new DOT system is equivalent to the old DOT system by expectation value.

Observe that $m_{k'}\in O(n^{k'})$, and introduce a collection of numbers
$\bar{n}(k')=\lceil (m_{k'})^{1/k'}\rceil$, and let $\bar{n}:=\max_{k'}\bar{n}(k')$.
Then, we can substitute the upper bounds $m_{k'}$ of the sums in the Hamiltonian $\mathbb{H}$
by $\bar{n}^{k'}$, such that all additional elements in the new Hamiltonian $\mathbb{\bar{H}}$
is zero in order to make sure the Hamiltonian does not change, physically.

The new Hamiltonian $\mathbb{\bar{H}}$ takes the form
\begin{align}\label{}
  \mathbb{\bar{H}}&=\sum_{l_1}^{\bar{n}} \sum_{i_1,j_1}^d \hat{\psi}_{l_1;i_1}^\dagger \bar{H}_{l_1;i_1 j_1}^{[1]}\hat{\psi}_{l_1;j_1}
  +\sum_{l_2}^{\bar{n}^2}\sum_{i_1,i_2,j_1,j_2}^{d}\hat{\psi}_{l_2;i_1i_2}^\dagger\bar{H}^{[2]}_{l_2;i_1i_2 j_1j_2}\hat{\psi}_{l_2;j_1j_2}
  +\cdots  \\ \nonumber
  &+\sum_{l_k}^{\bar{n}^k}\sum_{i_1\cdots i_k,j_1\cdots j_k}^{d}\hat{\psi}_{l_k;i_1i_2\cdots i_k}^\dagger \bar{H}^{[k]}_{l_k;i_1\cdots i_k j_1\cdots j_k} \hat{\psi}_{l_k;j_1j_2\cdots j_k}.
\end{align}

Combine the indices $l_k$ with $i_1i_2\cdots i_k$ and $j_1j_2\cdots j_k$ together,
and then each of the indices is upper bounded by $d\bar{n}$.
The Hamiltonian becomes
\begin{align}\label{}
  \bar{\mathbb{H}}&=\sum_{i_1,j_1}^{d\bar{n}} \hat{\psi}_{i_1}^\dagger \bar{H}_{i_1j_1}^{[1]}\hat{\psi}_{j_1}
  +\sum_{i_1,i_2,j_1,j_2}^{d\bar{n}}\hat{\psi}_{i_1i_2}^\dagger \bar{H}^{[2]}_{i_1i_2j_1j_2}\hat{\psi}_{j_1j_2}
  +\cdots \\ \nonumber
  &+\sum_{i_1\cdots i_k,j_1\cdots j_k}^{d\bar{n}}\hat{\psi}_{i_1i_2\cdots i_k}^\dagger \bar{H}^{[k]}_{i_1\cdots i_k j_1\cdots j_k}\hat{\psi}_{j_1j_2\cdots j_k}.
\end{align}

This is the second quantized form of a Hamiltonian $\hat{\bar{H}}$ of $k$ interacting systems each of which is $d\bar{n}$-dimensional
\begin{equation}\label{}
  \hat{\bar{H}}=\hat{\bar{H}}^{[1]}+\hat{\bar{H}}^{[2]}+\cdots +\hat{\bar{H}}^{[k]},
\end{equation}
and $\hat{\bar{H}}^{[k]}$ is the operator for $\bar{H}^{[k]}_{i_1\cdots i_kj_1\cdots j_k}$.
$\hat{\bar{H}}$ acts on a Hilbert space $\mathcal{\bar{H}}$ with dimension $(d\bar{n})^k$.
In this form, the total dimension depends on the locality of interaction clearly,
and particularly, the dimension does not scale exponentially with $n$,
the number of the original qudit particles.
Since $m_{k'}\in O(n^{k'})$, we have $\bar{n}\in O(n)$, which leads to the
Hilbert space dimension in order $O((dn)^k)$.

The new dynamics $(\hat{\bar{H}}, \mathcal{\bar{H}}, |\bar{\psi}\rangle)$
for (not normalized) state $|\bar{\psi}\rangle\in\mathcal{\bar{H}}$
\begin{equation}\label{}
  |\bar{\psi}\rangle=\sum_{i_1i_2\cdots i_k}^{d\bar{n}} \psi_{i_1i_2\cdots i_k}|i_1,i_2,\cdots ,i_k\rangle
\end{equation}
is equivalent to the dynamics $(\mathbb{H},\mathcal{F},|\chi\rangle)$ based on the relation between first and second quantization.
As the result, the new DOT system is equivalent to the original DOT system
\begin{equation}\label{eq:energy}
\langle\psi|\hat{H}|\psi\rangle=\langle\bar{\psi}|\hat{\bar{H}}|\bar{\psi}\rangle.
\end{equation}
This indicates that a many-body system in first-quantized form can be in the first place be modeled in space $\mathcal{\bar{H}}$
which contains efficiently number of quantum states rather than $\mathcal{H}$ which has a vast inaccessible volume.

The state $|\bar{\psi}\rangle$ is a direct sum of $\bar{n}^k$ local states $|\bar{\psi}_j\rangle$ each of $d^k$ dimension.
As the local states are from a global state $|\psi\rangle$ in the original DOT system,
the set of local states $\{|\bar{\psi}_j\rangle\}$ are constrained
such that there exists a consistent global state $|\psi\rangle$.

In addition, there is a simpler way to introduce the new Hilbert space.
Each term in $\hat{H}$ can be viewed as a $k$-local term,
so there are totally $m$ $k$-local terms.
Then it is direct to obtain that
the new Hilbert space $\mathcal{\bar{H}}$ has dimension $d^k m$ in the order $O((dn)^k)$.

Roughly, the physical picture is that, given one $n$-qudit system, a $dn$-dimensional Hilbert space is formed,
which can be viewed as the direct sum of the individual Hilbert spaces for each qudit.
Then, tensor product structure is formed according to the interaction,
such that the interactions among the qudits are translated to the interactions among the $dn$-dimensional particles.
This looks similar with the process of forming a classical scale field from a collection of particles.
However, there is no tensor-product structure due to interaction in the model of classical scale field.
In other words, our model can be viewed as an ``eclecticism'' of classical and quantum models for many-body system,
in that there is only direct-sum structure in the classical model,
while there is only tensor-product structure in the standard quantum model.

The phase space of a classical many-body system is the direct sum of each phase space
of the local particles.
The reason is that we do not consider a global state on the system,
or say, the global state is always a product state.
A quantum system in Hilbert space can be converted into a phase space,
the number of variables is exponential as $d^n$.
If we reduce it to $d^km$,
the field variables become dependent with each other,
and it becomes a constrained Hamiltonian dynamics system.

\begin{example}
Quantum Heisenberg model.

\emph{Consider quantum Heisenberg model in arbitrary spatial dimension $(D=1,2,3)$
with at most 2-body interaction and arbitrary interaction geometry.
The Hamiltonian can be written as
\begin{equation}\label{}
  \hat{H}=\sum_{i=1}^n \hat{H}_i+\sum_{i,j}^m \hat{V}_{ij}.
\end{equation}
The 2-body interaction $\hat{V}_{ij}$ can be nearest-neighbor or not.
$m\leq n^2$.
From the procedure to construct the eclectic model,
we find the energy is
\begin{equation}\label{}
  H=\sum_{l_1}^n \sum_{ij}^{d} \psi^*_{l_1;i}H_{l_1;ij} \psi_{l_1;j} +\sum_{l_2}^m \sum_{ijkl}^{d} \psi^*_{l_2;ij}V_{l_2;ijkl} \psi_{l_2;kl},
\end{equation}
for 2-body state $|\psi_{l_2}\rangle=\sum_{ij}^{d} \psi_{l_2;ij} |ij\rangle$.}

\emph{Since there are only $m$ 2-local terms, we need to add zero elements to construct a $4n^2\times 4n^2$ matrix $\bar{V}$.
We can proceed and obtain the eclectic Hamiltonian
\begin{equation}\label{}
  \hat{\bar{H}}=\sum_{ij}^{dn} H_{ij}|i\rangle\langle j| +\sum_{ijkl}^{dn} \bar{V}_{ijkl} |i,j\rangle\langle k,l|
  \equiv \hat{H}^{[1]} + \bar{V},
\end{equation}
and eclectic state $|\bar{\psi}\rangle=\sum_{l_2}^{n^2} |l_2\rangle|\psi_{l_2}\rangle$
such that $H=\langle \bar{\psi}|\hat{\bar{H}} |\bar{\psi}\rangle$.
The eclectic system contains two interacting $dn$-level systems,
the total dimension is $(dn)^2$ instead of $d^n$.
The property of the system is determined by the set $\{|\psi_{l_2}\rangle\}$.}
\hfill $\blacksquare$
\end{example}

\subsection{Relations with other approaches}

\subsubsection{Slave-particle model}

According to spin-charge separation~\cite{Hal81},
there could be quasiparticles called spinon and chargon corresponding to the spin and charge
degree of freedoms of a particle, e.g. electron.
Since electron is fermion, the spinon could be fermion if the chargon is boson,
or vice versa.
For a many-body spin system, the spin operator is substituted by field operators satisfying some constraints.
For instance, in Schwinger boson model for spin-1/2 particle~\cite{AA08},
set $\sigma^+=\hat{a}^\dagger \hat{b}$, $\sigma^-=\hat{b}^\dagger \hat{a}$
for two bosonic fields $\hat{a}$ and $\hat{b}$.
Since spin operator satisfies  $[\sigma^+, \sigma^-]=2\sigma^z$, then
$2\sigma^z=\hat{a}^\dagger \hat{a}-\hat{b}^\dagger \hat{b}$.
For the number of fermions, it is required that $\hat{a}^\dagger \hat{a}+\hat{b}^\dagger \hat{b}=2S$,
and $S$ is the spin magnitude operator.
The Hamiltonian involving spin operators $\hat{H}_{\text{spin}}$ is modified to a new Hamiltonian involving
spinon fields $\hat{H}_{\text{spinon}}$ along with constraints.
The quantum state $|\psi_{\text{spin}}\rangle$ in Hilbert space $\mathcal{H}$
is substituted by quantum state $|\psi_{\text{spinon}}\rangle$ respecting statistics in Fock space $\mathcal{F}$
along with constraints.
The dynamics in $\mathcal{F}$ is only restricted or projected in a subspace corresponding to dynamics in $\mathcal{H}$~\cite{Wenbook,Wen02}.
However, the constrained dynamics in $\mathcal{F}$ is no easier to solve than the original dynamics in $\mathcal{H}$.

The slave-particle model is different from the method based on quantization hierarchy.
In slave-particle model, we see that the first-quantized spin operator is set equal to second-quantized field operator.
In our approach, the field operator in the second-quantized model of a first-quantized many-body system, e.g. spin system,
does not have to satisfy constraints except the statistics,
since the field operators are not equal to the first-quantized operators in any sense.
For instance, for spin-1/2 particle, the Pauli operators $\sigma^i$ ($i=x,y,z$) is second-quantized to
$\hat{\psi}^\dagger\sigma^i\hat{\psi}$ for fermionic field $\hat{\psi}$.
The algebra of operators $\sigma^i$ and algebra of field $\hat{\psi}$ are two different algebras,
instead of the same algebra as in the slave-particle model.
In other words, the slave-particle model aims to solve a first-quantized system with field theories
such that the solution is expected to be valid in the first-quantized space;
however, our method is to convert a first-quantized system into a second-quantized one
such that the solution in the new second-quantized space is equivalent to a corresponding solution in the
first-quantized space.

\subsubsection{Jordan-Wigner transformation}

The Jordan-Wigner transformation~\cite{JW28} and related forms~\cite{BK02} also follow the spirit that to make a direct equivalence between spin operator and fermion.
This approach is useful for converting a spin system to a fermion system and vice versa,
and also can be employed in quantum simulations using quantum computers which works in first quantization~\cite{Llo96,Fey82}.

The equivalence of algebras between spin operator and field operator
demonstrated by slave-particle model or Jordan-Wigner transformation
is different from the equivalence of dynamics in the quantum hierarchy.
In fact, equivalence of algebras cannot make sure the dynamics are also equivalent
since it does not necessarily ensure the equivalence of algebras for {\em all} operators.
Furthermore, the quantum hierarchy provides an alternative which is more fundamental
and can make connections between first-quantized and second-quantized systems,
and no equivalence of algebras is required.

\section{Discussions}
\label{sec:dis}

In this work, we developed the operator-valued quantum phase space formalism.
The formalism generalizes the usual Geometric Quantum Mechanics
and the usual classical phase space (with position and momentum) to operator-valued versions,
and provides a natural link between first and second quantization.

The Boson/Fermion statistics (commutative/anticommutative commutation relation)
is derived merely according to Schr\"odinger equation and Heisenberg equation without referring to Relativity or spin.
One of the basic principle in quantum theory is the {\em identical particle postulate}.
From our study, the assumption of indistinguishability becomes ``phenomenological''
in that as long as the same degree of freedom of different particles are studied, e.g. the spin of electrons,
these particles can be treated identically such that they can only be bosonic or fermionic
(ignoring the other cases for composite particles),
which does not necessarily imply that we cannot detect the differences among them.
In other words, these particles are rather ``congeneric'' instead of identical.

The combination of quantization and hamiltonization yields the concept of quantization hierarchy.
We have discussed some basic properties of the hierarchy.
For a particle (or system, excitation),
its dynamics on different levels of the hierarchy are equivalent with each other
in the sense that the expectation values of dynamical variables are equivalent.
In fact, it is quite difficult to decide which level of its hierarchy a particle is on.
For instance, photon is described in Fock space.
It is possible that there exists a dynamics in Hilbert space
and a classical dynamics in some manifold of a certain particle,
which could be ``ether''.
However, the program of quantization hierarchy can be doubted
since there seems to be an infinite levels in the hierarchy.
The higher-level dynamics might be redundant
since there is no obvious evidence for its existence so far.

The quantization method is applied to quantum many-body dynamics.
Particularly, we showed that a local Hamiltonian dynamics can be formalized in a Hilbert space
whose dimension does not scale exponentially with the number of particles in the system.
For $k$-local interacting $n$-qudit system, the dimensional is about $(dn)^k$.
However, this does not imply \textsc{qma}-complete problem,
such as the \textsc{local hamiltonian} problem~\cite{KSV02} of finding the ground state and associated properties
can be solved efficiently.
In contrast, the operator-valued quantum phase space formula and the eclectic model
have more appealing potential for the study of few-body quantum dynamics,
such as those in quantum control and quantum chaos.

\end{spacing}
\bibliography{phaspace}
\end{document}